\documentclass[preprint,showpacs]{revtex4}
\usepackage{bm}
\input{epsf}
\begin{document}
\title{Diffraction of light by interfering liquid surface waves}
\author{Tarun Kr. Barik, Anushree Roy and Sayan Kar}
\email{sayan@phy.iitkgp.ernet.in}
\affiliation{Department of Physics, Indian Institute of Technology,
Kharagpur 721 302, WB, India}


\begin{abstract}
Interfering liquid surface waves are generated by electrically driven
vertical oscillations of two or more equispaced pins immersed in a liquid
(water). The corresponding intensity distribution, resulting from
diffraction of monochromatic light by the reflection phase grating formed
on the liquid surface, is calculated theoretically and found to tally with
experiments. The curious features of the diffraction pattern and its
relation to the interference of waves on the liquid surface are used to
measure the amplitude and wavelength of the resultant surface wave along
the line joining the two sources of oscillation. Finally, a sample
diffraction pattern obtained by optically probing surface regions where
interference produces a lattice--like structure is demonstrated and
qualitatively explained.  
\end{abstract}
\maketitle
\def\d{{\mathrm{d}}}
Surface waves are ubiquitous in nature. Also known
as the Rayleigh wave, they appear on solid as well as liquid surfaces
resulting in diverse phenomena with useful applications. In particular, it
is well established that surface acoustic waves (henceforth SAW) are of
prime importance in solid state technology {\cite{morgan:2000}}.
Investigations on SAW in liquids has been relatively less.
The traditional `ripple tank' method of measuring the surface tension 
of a liquid is one of the earliest studies \cite{surface}. 
Recent advances 
along related directions (thermally activated liquid surfaces, 
spatial damping of liquid SAW, light scattering/photon correlation 
spectroscopy of liquid SAW, to name a few) 
have been carried out and reported in  
\cite{earnshaw:1988, langevin:1992,lee:1993,kolevzon:1996, 
klipstein:1996}. The basic principle of
light diffraction by liquid SAW remains the same as in solids, though
their quantitative features, the generation mechanism surely
differ. Ripples on liquid surfaces, 
caused by external vibrations (usually electrically/thermally driven)
act as an effective dynamical phase grating for incident radiation.
The complex aperture function
of the phase grating is determined from the phase shift of the
reflected wave due to variation (sinusoidal) of the surface height 
\cite{duncan:2000}.
A Fourier transform of the aperture function gives the light field 
strength on the screen from which the intensity can be obtained.
Features of the intensity distribution provide information on
the liquid SAW as well as properties of the liquid itself. Since
liquid properties can be measured by other methods too, one is usually
more interested in finding out the nature of the SAW using external
non--destructive probes (such as light).  
Recently, Miao et al \cite{miao:2000} have studied diffraction of light
from circular ripples on water and experimentally established the
correlation between the diffraction pattern and the SAW amplitude.
Using the features of the intensity pattern, they have also measured
the SAW wavelength.
In this letter, we focus our attention on
diffraction by {\em interfering} SAW, a topic which does not seem to have
been dealt with much in the past. In other words, we
generate superposed waves on the liquid surface by electrically driven
vertical oscillations of slightly immersed metallic pins. 
The resulting interference
pattern on the liquid surface diffracts incident light. Our aim is to
first investigate how the {\em superposed} character of the
wave pattern on the liquid surface
can be {\em seen} in the diffraction intensity. 

\begin{figure}
\centerline{\epsfxsize=3.00in\epsffile{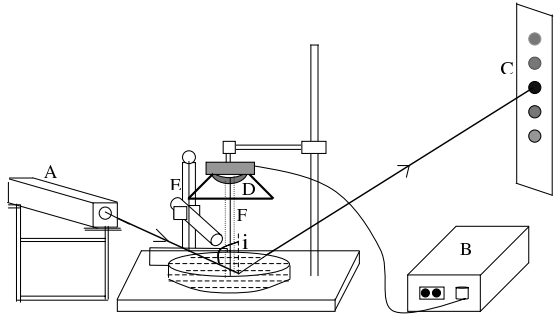}}
\caption{Schematic diagram of the experimental setup to study SAW on
liquid, A: Laser, B: Frequency generator, C: Screen, D:
Loudspeaker, E: Microscope, F: Exciter}
\label{Figsaw1}
\end{figure}

\begin{figure}
\centerline{\epsfxsize=3.00in\epsffile{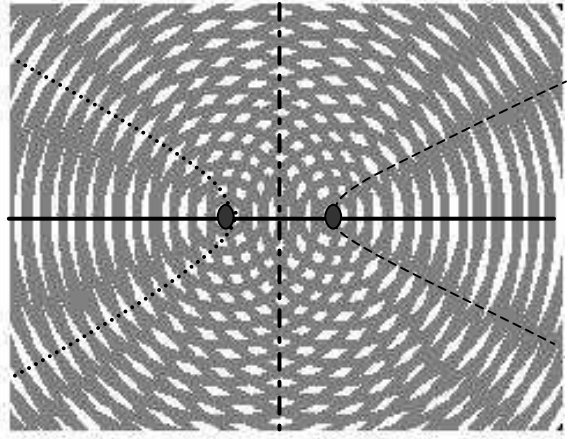}}
\caption{Line drawing of interference between two circular waves. Dotted, 
dashed and 
dashed-dotted lines are explained in the text}
\label{Figsaw2}
\end{figure}

A schematic diagram of the experimental set up is shown in 
Fig.~\ref{Figsaw1}. 
Water (about 1 cm deep) is kept in a petri-dish of diameter 18.5 cm. 
A pin is glued to the diaphragm of a small loud 
speaker,
which is driven by a low frequency signal generator.
A sinusoidal signal of frequency $\Omega$ (220 Hz in our experiment) is
used to make the pin vibrate, which acts as a SAW exciter.
Best results are obtained when the pin is just below the liquid surface.
If the pin is immersed more, splashing and other associated effects can
occur. To create superposed waves on the surface we have glued two pins on
the diaphragm of the same loudspeaker. The separation betweem the pins 
is measured by a travelling microscope.
The two pins oscillate simultaneously, in phase,
on the water surface and  create individual circular waves about their
location which, in turn, interfere to produce the well--known two--source
interference pattern, shown in Fig.~\ref{Figsaw2}.
A 5 mW He-Ne laser of wavelength $\lambda$ (632.8 nm)
and of  beam-diameter 1.5 mm is directed to fall on the water surface,
where the superposed SAW is formed. The angle of illumination (i.e. the
angle at which the laser beam falls on the liquid surface), i, in our
experiment is 77 degrees.
It is known that at  frequencies of the order of 220 Hz
the SAW decay length is much greater than the size
of the illuminated spot \cite{kinsler:1982}. Due to Fraunhofer diffraction
of light by the SAW
phase grating, spots are observed on a screen, placed at a reasonably
large distance (3.15
meters in our case) from the water surface. The images of the
diffraction pattern are taken using a digital camera. The spurious noise 
in the images is removed by using the `Photoshop' software.  The 
intensity of the diffraction spots have been measured by a photodiode detector. 
The linear relation between the photodiode current and the intensity of 
diffracted light has been checked.

The circular waves generated on the surface of the liquid due to more than
one source of disturbance (oscillation) can be approximated as a
linear superposition of individual sinusoidal waves. We can write the
individual waves as $\psi_1 = h \sin \left  (\Omega t - k r_1 \right)$
and $\psi_2 = h \sin \left (\Omega t - k r_2\right )$, where we assume
that the two oscillations have the same frequency $\Omega$, wavenumber $k$
and amplitude $h$. $r_1$ and $r_2$ are the distances of the oscillation
sources from the point of observation.
In addition, we would be interested in focusing light at a point
along the line joining the two sources (solid line in 
Fig.~\ref{Figsaw2}).
A sum of these two waves produces a resultant wave profile (along the
line joining the sources):

\begin{equation}
\psi = 2 h \cos {\frac{kd}{2}} \sin \left (\Omega t - k r \right )
\label{eqn1}
\end{equation}

\noindent
where we have assumed $r_2-r_1=d$, $r_2+r_1=2r$.
In general, for a superposition of $N$ such oscillations we expect the
factor $2 \cos \frac{kd}{2}$ to be replaced by $\frac{\sin \frac{Nkd}{2}}
{\sin \frac{kd}{2}}$ (reminiscent of the factor in the diffraction formula 
for $N$ slits
in the theory of Fraunhofer diffraction)\cite{hecht:2003}. 
Thus, the resultant 
phase modulation function for the reflected light from 
the superposed liquid surface waves turns out to be \cite{duncan:2000} :

\begin{equation}
\phi (x) = \frac{2\pi}{\lambda} \left [ (4 h \cos \frac{kd}{2} \cos i)
\sin \left ( \Omega t - \frac{kx}{\cos i} \right ) \right ].
\label{eqn2}
\end{equation}
\noindent
We can easily rewrite this formula with an equivalent `$h$' to make it look
like the phase modulation by a single wave. This will yield the
replacement $h'=2 h \cos \frac{kd}{2}$. 
In general for $N$ sources of
oscillation :
$h'= h\frac{sin \left ( Nkd/2\right )}{sin \left (kd/2\right )}$.
The light field strength ($E$) is the Fourier transform of the object
(aperture) function given by $\exp \left (j \phi\right )$.
The intensity is obtained by considering $E E^*$ and
yields \cite{goodman:1968}:

\begin{equation}
I(x') = \sum_n J_n^2 \left (4\pi h' \cos i/ \lambda \right )
\delta \left (\frac{x'}{\lambda z} - \frac{n}{\Lambda \cos i}\right )
\label{eqn4}
\end{equation}

\noindent
where $z$ is the horizontal distance between the location of the
laser spot on the liquid surface and the screen, SAW wavelength $\Lambda=
\frac{2\pi}{k}$. $x'$ is the position variable on the observation plane.

\begin{figure}
\centerline{\epsfxsize=3.00in\epsffile{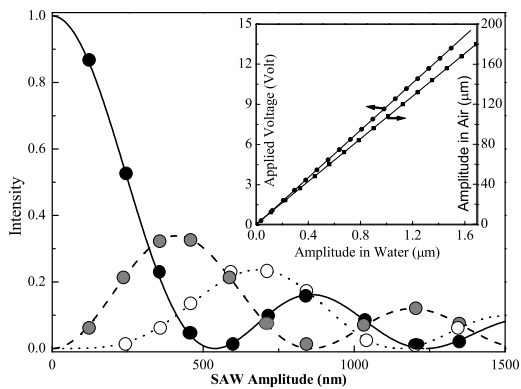}}
\caption{Theoretical plot of intensity vs. SAW amplitude for
zero order (solid line), first order (dashed line) and second order
(dotted line) diffraction spots using eq. ~\ref{eqn4}. Experimentally 
measured
intensity for these spots are shown by black filled, grey filled and open
circles, respectively. The linearity in variation of resultant SAW
amplitude in water and applied voltage in signal generator (also
amplitude imparted by the exciter) is shown in the inset}
\label{Figsaw3}
\end{figure}

The intensity distribution, given in eq. ~\ref{eqn4}, demonstrates that 
at
the location of zeros of the Bessel function one would get vanishing
intensity of the
different orders at different SAW amplitude. For instance, for $n=0$,
and $x'=0$, the intensity will
vary as $J_0^2$. Thus, it will vanish at the value of $h$ for which
$J_{0}^{2}$ has a
value zero. Interestingly, for the superposed waves the value of $h$ for
which the intensity zero will be observed is a
factor $\frac{1}{2\cos \frac{kd}{2}}$ (for $N=2$)
smaller than the value for a single
wave. We find that the intensity zero at
zeroth order will appear nearly at half the $h$ value obtained for
the single wave case when $\left\vert \cos(\frac{kd}{2})\right\vert 
\rightarrow 
1$.
In our experiment, since $h$ is proportional to the voltage, we should
be able to see this reduction in $h$ through the reduction in the applied 
voltage of the frequency generator.
For more than two sources of oscillation
placed close by, we would expect a reduction by a factor of $N$ (since
$\vert\sin\frac{Nkd}{2}/\sin \frac{kd}{2}\vert$
tends to $N$). 

Fig.~\ref{Figsaw3} shows a plot of the intensity as a function of the SAW 
height $h$. The black filled, grey filled and open circles are
experimentally measured data points for zero, first and second order
diffraction spots, respectively; which  verifies
the well-established theoretical curve, obtained from eq. ~\ref{eqn4}.
In the inset of this figure we have shown the
relation between the applied voltage in the function generator
(amplitude of oscillation imparted to the pin)
and the amplitude of the resultant SAW. A linear correlation is
seen though we note that the SAW amplitude is approximately 100 times
smaller than the oscillation amplitude of the exciter. We
mention that we have obtained this correlation by first calculating the SAW
amplitude from eq. ~\ref{eqn4} and then relating it with the applied 
voltage
for the same intensity on the diffraction pattern. We have
measured the imparted amplitude of vibration of the pin (in air)
for different values of the applied voltage, using a travelling 
microscope.

\begin{figure}
\centerline{\epsfxsize=3.00in\epsffile{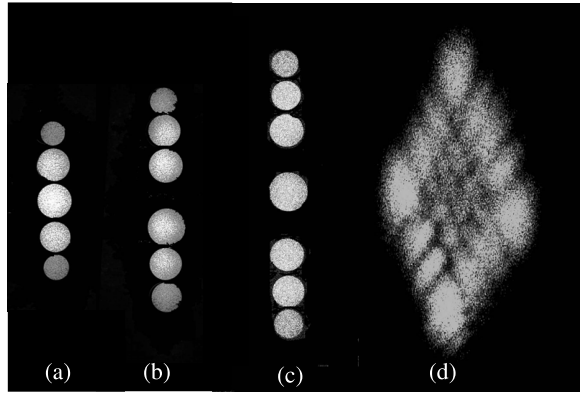}}
\caption{Diffraction pattern of SAW on water surface}
\label{Figsaw4}
\end{figure}

Fig.~\ref{Figsaw4}(a) shows the Fraunhofer diffraction patterns of 
monochromatic light
by SAW when a single pin has been used as SAW exciter. As the loudspeaker
amplitude (applied voltage) is increased slowly,
at a certain amplitude of vibration the central spot vanishes (Fig.
4(b)). We observe the reappearance(disappearance) of the central 
spot(first order spot) when the amplitude is
increased further (Fig.~\ref{Figsaw4}(c)). For a single pin this vanishing 
of the
central and first order diffraction spots takes place
for applied voltages 4.5 V and 7.4 V, respectively. Exactly similar
diffraction patterns appear for two superposed waves (with light incident along
the line joining the sources)
for applied voltages 2.4 V and
3.9 V, respectively. In other words, for two pins zero order diffraction
spot vanishes for nearly half of the applied voltage compared to that in 
single
pin, as we expect from the above theoretical understanding.
Similarly, for three pins the zero order spot vanishes when the applied
voltage is 1.7 V.

The displacement (x$^{\prime}$) of the nth order spot from the centre of
the diffraction pattern is $\frac{n\lambda z}{\Lambda cos i}$.
Measuring $x'$ for single pin, we determined
the SAW wavelength in our experiment to be 2.1 mm.
Moreover, with more than one pin (say two)
for certain special values of $kd/2$ two interesting
possibilities emerge,
which, in turn can be used to check the wavelength of the superposed
liquid surface wave along the line joining the sources.

\noindent
{\em Complete destructive interference}:
It is easy to note that if $d= \frac{m \Lambda}{N}$, where $m$ is an
integer but {\em not} equal to or {\em not} a multiple of $N$ the factor
$\sin\frac{Nkd}{2}/\sin \frac{kd}{2}$
vanishes. If $N=2$, this factor
vanishes for $d=\frac{\Lambda}{2}, \frac{3\Lambda}{2}, \frac{5 \Lambda}{2}
....$. Similarly, for other values of $N$. This in turn implies that
for these values of $d$ the argument of the Bessel function would vanish.
But $J_n (0)$ is non--zero
only for $J_0$, which implies that we see only a central spot at maximum
(100 $\%$) diffraction efficiency. This special feature, essentially
arising out of complete destructive interference can very easily
check the value of the wavelength of the liquid surface waves.
Keeping in mind that the change in separation between the two pins ($N$=2) 
does not change the SAW wavelength estimated above, this should correspond 
to $d = \Lambda/2=1.05$ mm, $3 \Lambda /2=3.15$ mm and $5\Lambda/2=5.25$ 
mm.   
In our experiment, when the separations between the two pins are 1.0 mm, 
3.0 mm and 5.2 mm except the central order, the intensities of the other 
diffraction spots go to zero irrespective of SAW amplitude.

\noindent
{\em No change in diffraction pattern with changing N and h}:
A second feature emerges when the diffraction pattern
remains unchanged even when we increase the number of oscillation sources
(pins). This will happen when the factor
$\sin\frac{Nkd}{2}/\sin \frac{kd}{2}$
takes on the value one, or when $d= \frac{m\Lambda}{N+1}$, where
$m$ is an integer but {\em not} equal to or {\em not} a multiple of $N+1$.
In particular, for $N=2$ we have
$d=\frac{\Lambda}{3},\frac{2\Lambda}{3}, \frac{4\Lambda}{3} ....$.
For $\Lambda$=2.1 mm, this observation must match
the values of $d= \Lambda/3=0.7 $ mm, $2\Lambda/3=1.4$ mm, 
$4\Lambda/3=2.8$ mm and  $ 5\Lambda/2=3.5$ mm.
In our experiment, when the separation between the two pins are 1.4 mm, 
2.7 mm and 3.4 mm we observe identical diffraction patterns as shown in 
Fig.~\ref{Figsaw4}. The zero order
diffraction spot vanishes for the same applied voltage 4.5 V in each case. With 
identical conditions, when we lift one of the two pins, we get the
vanishing of the central spot at the same voltage. We could not set the
separation between the pins to values smaller than 1 mm. This restricted
us from checking this phenomenon for lower values of d.
Therefore, the wavelength measured by
explicitly working with the argument of the delta function in the
intensity expression, can now be checked using superposed liquid
surface waves.

The dispersion relation [$\Omega^{2}={\alpha} k^{3}/\rho$, where $\alpha$
and $\rho$ are the surface tension and the density of the liquid,
respectively] also provides
a third (theoretical) check on the value of the SAW wavelength.
Using this relation we have calculated the SAW wavelength to be
2.1 mm. Finally, using the values
obtained above one can find out the value of the phase velocity
[$v=\sqrt{\frac{2\pi\alpha}{\Lambda\rho}}$] of the
liquid surface waves along the line joining the two sources
\cite{levich:1962}. We have
calculated $v_p$= 46 cm/sec.

It is certainly true that all our quantitative
results are for the region
{\em along the line joining the sources}.
At other oblique points of incidence (of laser light
on the surface), or more importantly in the
region where the interference pattern shows interesting
features, we have not been able to find the exact
analytical expression for the intensity distribution of diffracted light.
Fig.~\ref{Figsaw4}(d) shows the diffraction intensity from the liquid
surface when
the light is incident on a region near the dashed-dotted line in
Fig.~\ref{Figsaw2}.
The interference pattern on the liquid surface in this region shows a
`lattice--like structure' (more precisely a phase lattice). This
structure is indeed reflected in the corresponding diffraction pattern
shown in Fig.~\ref{Figsaw4}(d).
Further theoretical and experimental understanding of diffraction of
light
from this region is left for future investigation.

To conclude, we have shown how the diffraction of light
by interfering liquid surface waves can help us quantify the nature of
the superposed waves and also find their wavelength
along specific directions (here, the line joining the oscilating pins).
The change in the values of the liquid SAW height (amplitude) with the
number of oscillating sources is a measure of the "interference" effect
on the liquid surface. It is possible, 
therefore,
to comment on the number of oscillating sources by systematically observing
the changes in the diffraction pattern as a function of height (voltage).
Additionally, using the characteristics of the superposed SAW and the
resulting light diffraction features we have been able to provide
a way of crosschecking the SAW wavelength. Finally, we have 
been able to quantify how much wave amplitude is generated in the liquid
if the pins are vibrated at some definite amplitude of the exciter. This 
of course is liquid specific and is an entirely empirical consequence.

AR and TKB thank DST for financial support. The authors thank G. P. Sastry
for useful discussions and C. S. Kumar for his help with the photographs.



\begin{thebibliography}{0} 
\bibitem{morgan:2000} Morgan D. P., {\em Int.
Jr. of high speed electronics and systems}, {\bf 10} (2000) 553; and the
references therein. 
\bibitem{surface} Watson F. R., {\em Phys. Rev.} {\bf
12} (1901) 257. 
\bibitem{earnshaw:1988} Earnshaw J. C. and McCoo E,, {\em
Phys. Rev. Lett.}, {\bf 72}, (1994) 84. 
\bibitem{langevin:1992}{\em Light
scattering by liquid surfaces and complementary techniques}, edited by
Langevin D. (New York: Marcel Dekker) 1992. 
\bibitem{lee:1993}Lee K. Y.,
Chou T., Chung D.S. and Mazur E., {\em J. Phys. Chem.}, {\bf 97} (1993)
12876. 
\bibitem{kolevzon:1996} V. Kolevzon, G. Gerbeth and G. Pozdniakov, 
{\em Phys. Rev.} {\bf E 55}, (1997) 3134.
\bibitem{klipstein:1996}Klipstein W.M., Radnich J.S. and Lamoreaux S.K.,
{\em Am. J. Phys.} {\bf 64}(6) (1996) 758.
\bibitem{duncan:2000}Duncan B.D., {\em Appl. Opt.}, {\bf 39} (2000)
2888.
\bibitem{miao:2000}Miao R., Yang Z., Zhu J. and Shen C., {\em Appl. Phys.
Lett.}, {\bf 80} (2002) 3033.
\bibitem{kinsler:1982}Kinsler L.E., Frey A.R., Coppens A.B.
and Sanders J.V., {\em Fundamentals of acoustics}, 3rd Ed. (Wiley, New
york), 1982, pp.  141-162.
\bibitem{hecht:2003} Hecht E., {\em Optics}, 4th Ed. (Pearson Education
) 2003,p. 461.
\bibitem{goodman:1968} Goodman J.W.,{\em Introduction to Fourier optics},
(McGrew-Hill, San Francisco), 1968, p.62.
\bibitem{levich:1962}Levich V.G., {\em Physicochemical hydrodynamics},
(Prentice -Hall Inc.) 1962, p. 596.
\end{thebibliography}
\end{document}